\documentclass[12pt,authoryear]{article}
\usepackage{geometry}
\usepackage{amssymb}
\usepackage{enumitem}
\usepackage{lipsum, float}
\usepackage{lscape}
\usepackage{graphicx}
\usepackage{hhline}
\usepackage{natbib}
\usepackage{authblk, afterpage, adjustbox}
\usepackage{hyperref}
\usepackage{cleveref}

\author{John W. G. Addy}
\author{Jean Langhorne}
\affil{Malaria Immunology Laboratory,
		Francis Crick Institute,
		London,
		NW1 1AT,
		UK}
\title{A Proposed Method for Assessing Cluster Heterogeneity}

\begin{document}
\maketitle

\begin{abstract}
Assessing how adequate clusters fit a dataset and finding an optimum number of clusters is a difficult process. A membership matrix and the \textit{degree of membership} matrix is suggested to determine the homogeneity of a cluster fit. Maximisation of the ratio of the overall \textit{degree of membership} at cluster number lag 1 is also suggested as a method to optimise the number of clusters in a dataset. A threshold factor upon the \textit{degree of membership} is also suggested for homogeneous clusters. Cluster simulations were given to compare how well the proposed method compares against established methods. This method may be applied to the output of both hierarchical and k-means clustering.
\end{abstract}

\section{Introduction}
Optimising the number of clusters through the partitioning of a dataset is a difficult process, especially with a dataset of heterogeneous or nested clusters. Methods of optimising the number of clusters in multivariate analyses typically consider how well the cluster optimisation fits given the within-cluster-sum-of-squares ($W_k$), by observing an \textit{elbow} in $W_k$, as cluster number ($k$) increases. However, the $W_k$, or manipulations of it, may not consider the proximity of an observation to another cluster it has not been assigned to. Identifying this on the individual basis has been given and suggested through the use of fuzzy clustering \citep{Bezdek1974fuzzy, dunn1973fuzzy}, calculating the \textit{degree of closeness} of an observation $i$ belonging to cluster $k$. However, the membership of observation $i$ belonging to cluster $k$ may only inform about a single observation and not all observations within a cluster. Proposed is a method of determining the homogeneity of clusters given their membership. This method does not consider which type of clustering that has been used, but only the final cluster list and the associated distance matrix of all individuals.

An empirical estimation of where an \textit{elbow} occurs in the $W_k$ as cluster number increases is given by the GAP statistic and the 1-standard-error rule \citep{tibshirani2001estimating}. Other methods of optimising cluster number include the comparison of the between-cluster-sum-of-squares along with the $W_k$ \citep{calinski1974dendrite} and the average Silhouette method \citep{rousseeuw1987silhouettes}. However, by ignoring the heterogeneity of the overall cluster fit, through $W_k$, it may not be understood how well represented each cluster is given the overall membership of observations. The Silhouette method \citep{rousseeuw1987silhouettes} provides some information about cluster membership, by averaging over all calculated Silhouette values from the nearest neighbouring cluster. Once the homogeneity of clusters has been considered, this allows the identification of the \textit{degree of membership} of a \textit{cluster}, given the \textit{membership} of all clusters. The maximising of the $\phi$ ratio of such degrees of membership at cluster lag 1 is suggested as an estimation of the optimum number of clusters.

Demonstrated is the application of the proposed method on two datasets, followed by a simulation study, similar to \cite{tibshirani2001estimating}, showing the utility of the method compared to other established methods. A short review of other methods of optimising cluster classification has been given in an effort to show how this method may be used in conjunction with others methods. Hierarchical clustering was considered in this paper, although the methods proposed may be considered for other clustering methods.

\section{Distance Matrix and Membership}
Consider a dataset ($x_{ip}$) over $i$ observations ($i = 1, 2, ..., n$) and $p$ variates. Where the distance between each observation can be expressed as a squared distance matrix ($d^2_{ij}$), derived by the squared Euclidean distance between two observations
\[
d^2_{ij} = \sum_{p}(x_{ip} - x_{jp})^2.
\]
The aim of cluster analysis is to classify $x_{ip}$ into appropriate $k$ clusters, $C_1$, $C_2$, ..., $C_k$, by optimising a given a metric, such as the within-cluster-sum-of-squares ($W_k$).

The method proposed intends to identify the heterogeneity of a performed cluster analysis by considering the allocation of $x_{ip}$ to a cluster, $C_k$, given the observation membership ($M_1, M_2, ..., M_m$).\\
\\
The \textit{degree of closeness} ($\delta_{ik}$) may be considered as the closeness of point $x_i$ being within cluster $C_k$ given by,
\begin{equation}
\delta_{ik} = \frac{1/{\overline{d^2}_{\!\!ik}}}{\sum_{k}1/{\overline{d^2}_{\!\!ik}}}.
\end{equation}
Where, $\delta_{ik}$ (Equation 1) is calculated as the \textit{degree of closeness} of observation $x_i$ occurring in cluster $C_k$ by considering the mean distance of $x_i$ to all other points allocated to every cluster in $C_k$ in Euclidean space, $d_{ij}$. Resulting in,
\[
\overline{d^2}_{\!\!ik} = \frac{1}{|n_k|}\sum_{j\in k} d^2 _{ij}.
\]
Other fuzzy clustering algorithms consider a similar conditional membership, such as FANNY \citep{kaufman2009finding} and Fuzzy C-Means \citep{Bezdek1974fuzzy, dunn1973fuzzy}. The \textit{degree of closeness} of an observation to a cluster is therefore derived directly from the distance matrix.
\begin{equation}
\delta_{ik} = \frac{1/ \left( {\frac{1}{|n_k|}\sum_{j\in k} d^2 _{ij}} \right)}{\sum_{k}1/{\left( {\frac{1}{|n_k|}\sum_{j\in k} d^2 _{ij}} \right)}}
\end{equation}
From these, a consideration of a \textit{degree of closeness} for each observation to a cluster (Equation 1) may be given in Equation 2. \\
\\
Although informative about $x_i$ to $C_k$, $\delta_{ik}$ provides no insight into how well the data fits the clusters regarding overall membership or how many clusters is optimum. The membership ($m$) of each cluster is considered to be all observations ($x_i$) within each cluster $k$. A summation of all distances within a cluster is considered to form matrix $\gamma_{mk} = \sum_{i \in m}{\overline{d^2}_{\!\!ik}}$, over $x_i$, for membership $m$ and cluster $k$. Considerations of the distances of $x_i$ are no longer given, but of the related membership distance over all points to each cluster. In this instance $x_i$ may be considered as a parameter to be integrated over. 

The membership matrix, $1/\gamma_{mk}$, has dimensions of $m$-by-$k$. In the case of $k=1$, $\gamma_{mk}$ is a $1$-by-$1$ matrix, and the \textit{degree of membership} of a cluster occurring given its membership would therefore become 1. With $k = 1$ it is assumed there are no \textit{other} clusters to which membership may be ascribed ($m=1$). Therefore, this method of optimising may only be considered at $k>1$. In the case where singleton clusters appear (one observation to a cluster) $\gamma$ was set to 1 to prevent dividing by zero and being dominated by singletons.

By considering membership, the \textit{degree of closeness} regarding the overall membership of a cluster or that of another cluster may be obtained. The \textit{degree of membership} may be considered as
\begin{equation}
\delta_{mk} = \frac{1/\gamma_{mk}}{\sum_{k}1/\gamma_{mk} }
\end{equation}
with an overall measure of validity for each cluster $\delta_{\bullet k}$ is given below,
\[
\delta_{\bullet k} = \frac{\sum_{m}1/\gamma_{mk}}{\sum_{m}\sum_{k}1/\gamma_{mk}}
\]
and a measure of membership validity,
\[
\delta_{m\bullet} = \frac{\sum_{k}1/\gamma_{mk}}{\sum_{k}\sum_{m}1/\gamma_{mk}}.
\]
\\
$\delta_{m\bullet}$ in Equation 3, may be generalised to the distance matrix as,
\begin{equation}
\delta_{mk} = \frac{1/\left( \sum_{i \in k}\left( {\frac{1}{|n_k|}\sum_{j\in k} d^2 _{ij}} \right) \right) }{\sum_{k} 1/\left( \sum_{i \in k}\left( {\frac{1}{|n_k|}\sum_{j\in k} d^2 _{ij}} \right) \right)}.
\end{equation}

\section{Cluster Optimisation}
Considering the $\delta_{mk}$ matrix (Equations 3-4), marginalising over the clusters' membership within $\delta_{mk}$ where $k\equiv m$ may be used to obtain the degree of fit of cluster $C_k$ occurring given its own membership, $M_m$, over all clusters.
\[
\delta_{T}=\sum_{k\equiv m}\delta_{mk}\delta_{m\bullet}
\]
$\delta_{m\bullet}$ is given as above. Therefore, the degree of fit of the overall clusters being true given their membership, $\delta_{T}$, may be derived for every cluster number, $k$, as $k \subset \mathbb{N}$. $\delta_{T}$ may be given as a summary of homogeneity over all clusters, given their overall membership.

$\delta_{T}$ may be considered to be a method of identifying an adequacy of the estimated clusters to their respective individual membership. If $\delta_{T}$ was small, then the estimated clusters may be considered heterogeneous and therefore \textit{inadequate}. Similarly, as $\delta_{T} \rightarrow 1$, then the estimated clusters may be considered homogeneous and provide an \textit{adequate} fit to their membership.

The degree of fit  cluster membership may be represented in terms of ratios, in order to understand if there has been a large discrepancy in $\delta_{T}$ as the number of clusters increases. Where, the ratio of clustering at $k$ clusters are,
\[
\phi_{k} = \frac{(\delta_{T})_k}{1-(\delta_{T})_{k}}.
\]
Optimising the number of clusters by identifying an \textit{elbow} in the $\phi_{k}$ may be given by maximising the $\phi_{k}$ ratio at lag 1.
\begin{equation}
\Phi_1 = \frac{\phi_k}{\phi_{k+1}}
\end{equation}
Maximising the $\Phi_1$ in favour for cluster number $k$ against $k+1$ investigates how likely the clusters fit the data as $k$ increases. It should be noted if two or more maxima with similar estimates of $\Phi_1$ occur at lag 1, for separate cluster numbers, a direct $\Phi$ estimate and modification of Equation (5) may be considered.

\section{Thresholding on Degree of Membership}
$\delta_{mk}$ is a $m$-by-$k$ matrix, which forces to use all the distances within $\gamma_{mk}$. Therefore, the \textit{degree of membership} of a cluster is still estimated even if one or many clusters are homogeneous ($\delta_{mk^\neg} \rightarrow 0$). These small estimates from homogeneous clusters may influence the overall estimate of $\delta_{T}$ and $\phi$ ratio. This may be adjusted for by acknowledging a threshold upon $\delta_{mk^\neg}$ and set the level of $\delta_{mk^\neg}$ where it is believed there to be a more homogeneous cluster. For this study a threshold of $0.1$ is suggested, where any value of $\delta_{mk^\neg} < 0.1$ is assumed $0$.

\section{Other Cluster Number Optimising Methods}
The classical approach to determining the number of clusters is to consider within-cluster-sum-of-squares ($W_k$) as cluster number increases and choose a number of clusters where there appears to be an elbow.

This method has been discussed and built upon by \cite{tibshirani2001estimating} to derive the GAP statistic which considers the 1-standard-error rule by comparing the log of the pooled within-cluster-sum-of-squares against a reference distribution derived through multiple sampling. The computational implementation of the GAP statistic is given in \cite{tibshirani2001estimating}, with a short description given here. Consider the within-cluster-sum-of-squares, $W_k$, a comparison between the logarithm of $W_k$ against the expectation of $log(\overline{W}_{\!\!kb})$. Where, $\overline{W}_{\!\!kb}$ is the average of $b$ Bootstrapped samples.
\begin{equation}
GAP_k = \frac{1}{b}\sum_b log(\overline{W}_{\!\!kb}) - log(W_k)
\end{equation}
The optimum cluster number is then derived from a 1-standard-error rule, $GAP_k \ge GAP_{k+1} - s_{k+1}$. Where, $s_{k+1}$  is the standard error of the Monte Carlo estimates of $E\{log(\overline{W}_{kb})\}$. The method of determining a reference distribution is given as a uniform distribution or Principal Component rotating.\\
\\
Another method of cluster number optimisation includes one proposed by \cite{calinski1974dendrite}. This method compares the between-cluster-sum-of-squares ($B_k$) by $W_k$.
\begin{equation}
CH_k = \frac{B_k/(n-k)}{W_k/(k-1)}
\end{equation}
Although the GAP statistic and the \cite{calinski1974dendrite} method provide good ways to optimise the number of clusters, both only consider the adequacy of the cluster fit and not the membership of observations to their own respective clusters or to others. \\
\\
A method which considers the membership of clusters is the Silhouette statistic \citep{rousseeuw1987silhouettes}. The Silhouette statistic ($Sl_k$) considers the average distance between points within an observations cluster ($\overline{w}_{k}$) and the average distance between points in the nearest cluster ($\overline{b}_{k}$).
\begin{equation}
Sl_k = \frac{\overline{b}_{k}-\overline{w}_{k}}{max\{\overline{w}_{k}, \overline{b}_{k}\}}
\end{equation}
A comparison of the different methods given above are compared against the marginalisation of the \textit{degree of membership}, and the derivation of $\Phi_1$, at lag 1. This was given to compare which method may give the most appropriate clustering and where any potential biases occur. The GAP statistic was derived from the Cluster package \citep{cluster2018}, and the Silhouette and CH method from the ClusterCrit package \citep{clusterCrit2018}.

\section{Examples}
\subsection{Iris Datset}
The method of deriving a \textit{degree of membership} is applied to the petal width and length from the Iris dataset given by \cite{Anderson1935irises}. Complete linkage hierarchical clusters was used to cluster in this example.

The distribution of the petal width and length (Figure 1(a)) show overlap between clusters 2 ($\square$) and 3 ($\blacksquare$). Both GAP statistics estimate the optimum cluster number to be 4 (Figure 1(b-c)). The Silhouette method correctly estimates the number of 3 clusters (Figure 1(d) and the CH method over estimates the optimum cluster number with 6 (Figure 1(e)). The maximum $\phi$ ratio at lag 1 occurred at cluster number 3 and had a value of 2.61, with a \textit{degree of membership} at cluster 3 ($\delta_T$) of 0.940 (Figure 1(f)). Therefore, suggesting the clusters are not heterogeneous, regarding the membership fitted each cluster. Table 1 shows how clusters 2 (centres; 1.87, 5.29) and 3 (1.21, 3.96) are not that well defined compared to cluster 1 (0.25, 1.46). The $\delta_{11}$ was 0.985, which provides a good estimate that the membership of cluster 1 was represented by its own membership. $\delta_{22}$ and $\delta_{33}$ were 0.714 and 0.872, respectively, with $\delta_{23}$ estimated at 0.246 and $\delta_{32}$ estimated at 0.092. Therefore, the membership of cluster 2 did not fit cluster 2 as well as the membership of cluster 3 to cluster 3. The lack of representation of the membership of cluster 2 given its cluster is contributing the most to the overall $\delta_{T}$ at cluster 3. \\

\afterpage{
\begin{landscape}
\begin{figure}
\caption{\textbf{(a)} Summary of the Iris data \citep{Anderson1935irises} in example \textbf{6.1}, cluster 1 ($\circ$), cluster 2 ($\square$) and cluster 3 ($\blacksquare$). \textbf{(b)} estimated GAP, unif. \textbf{(c)} GAP, PCA. \textbf{(d)} Silhouette. \textbf{(e)} CH. \textbf{(f)} $\delta_{T}$ ($\bullet$, solid-line) and $\phi$ ratio at lag 1 ($\circ$, dashed-line).}
\includegraphics[width=5.5cm]{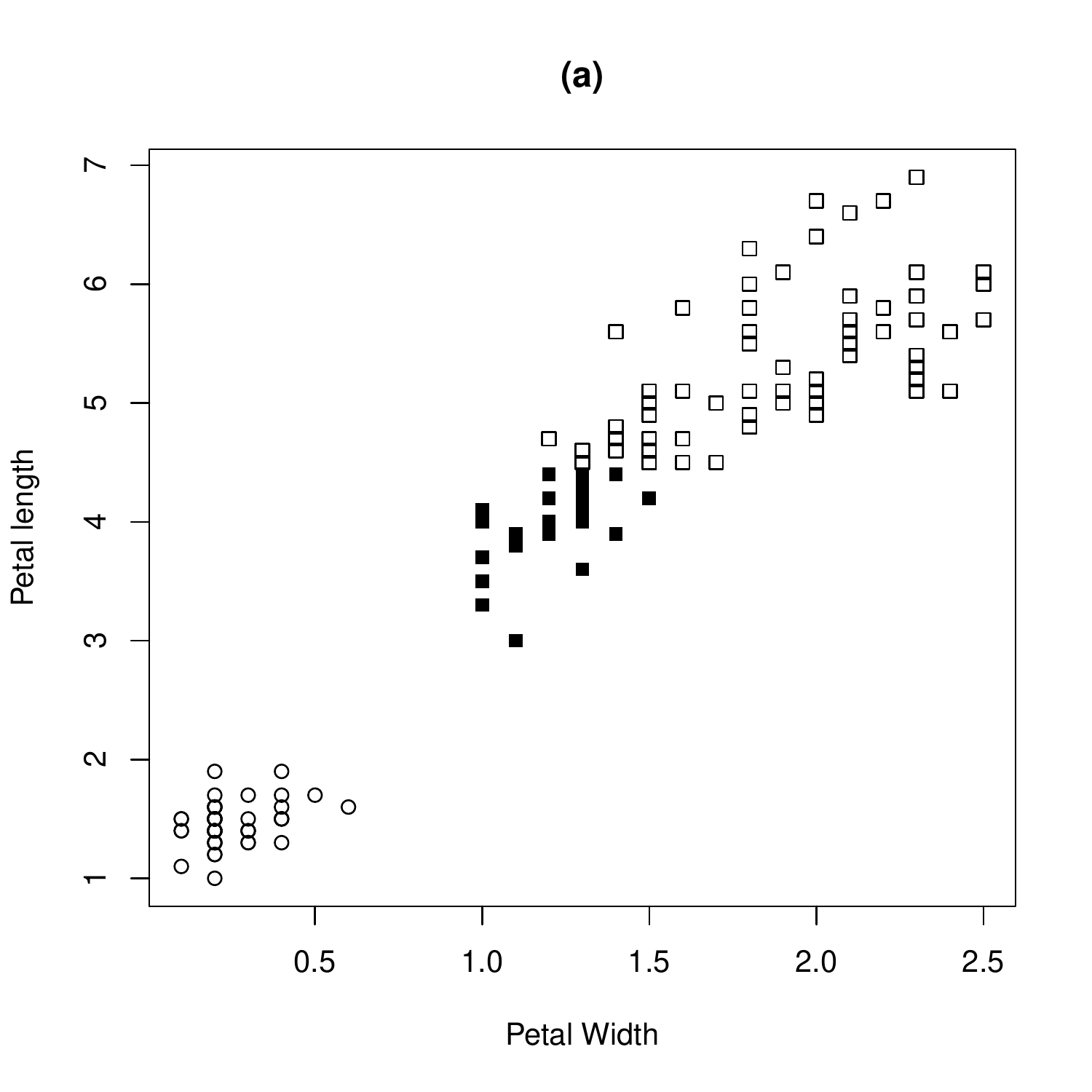}
\includegraphics[width=5.5cm]{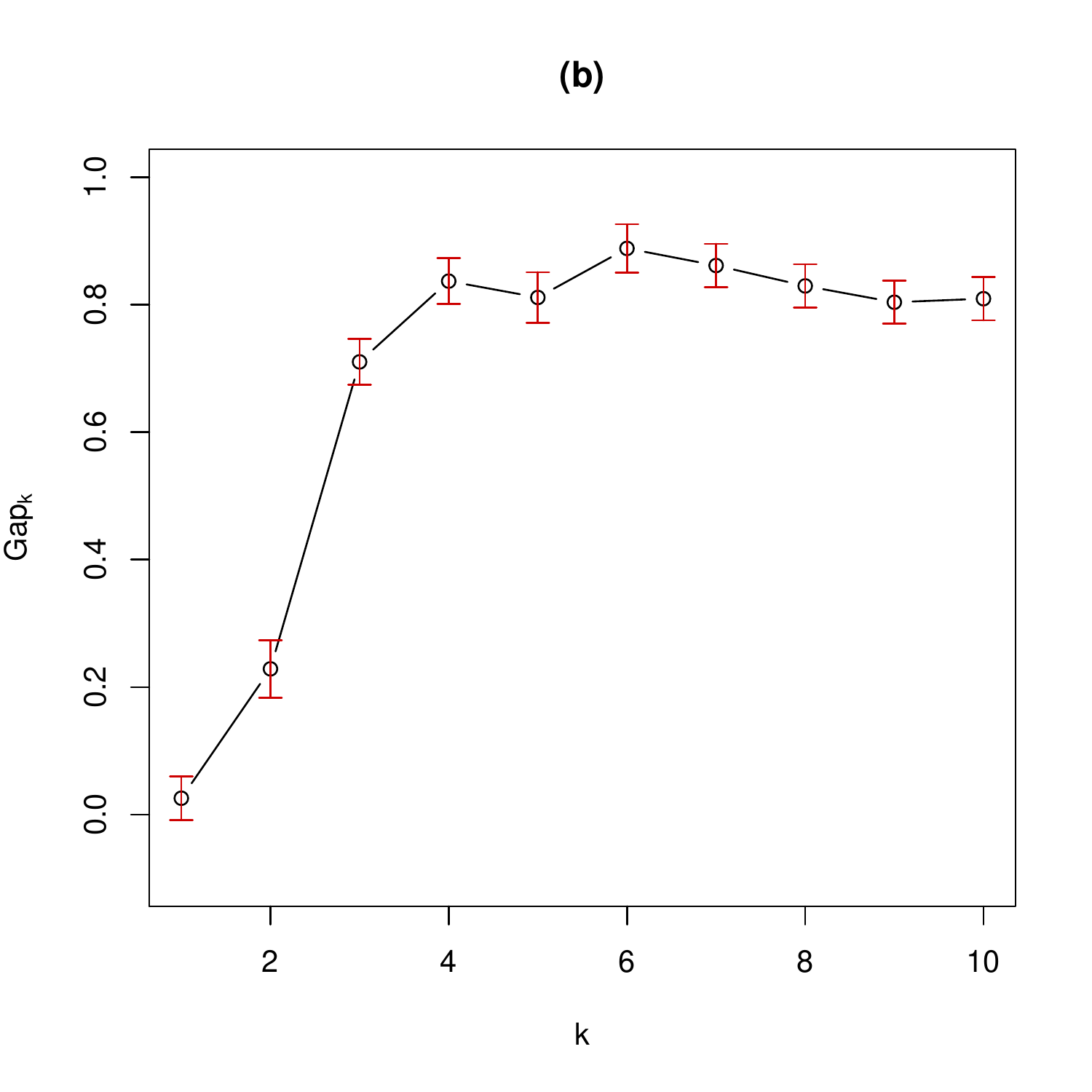}
\includegraphics[width=5.5cm]{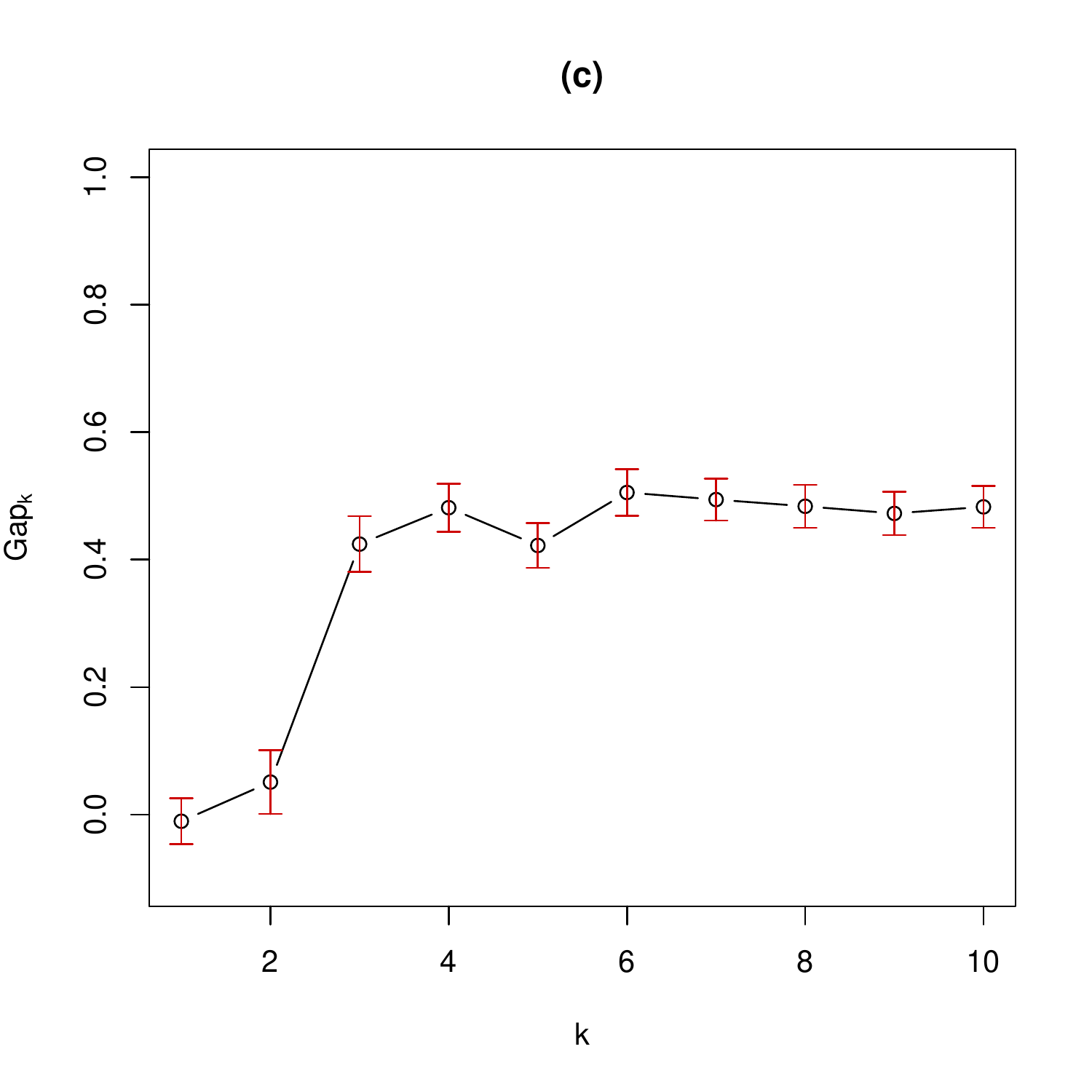}
\\
\includegraphics[width=5.5cm]{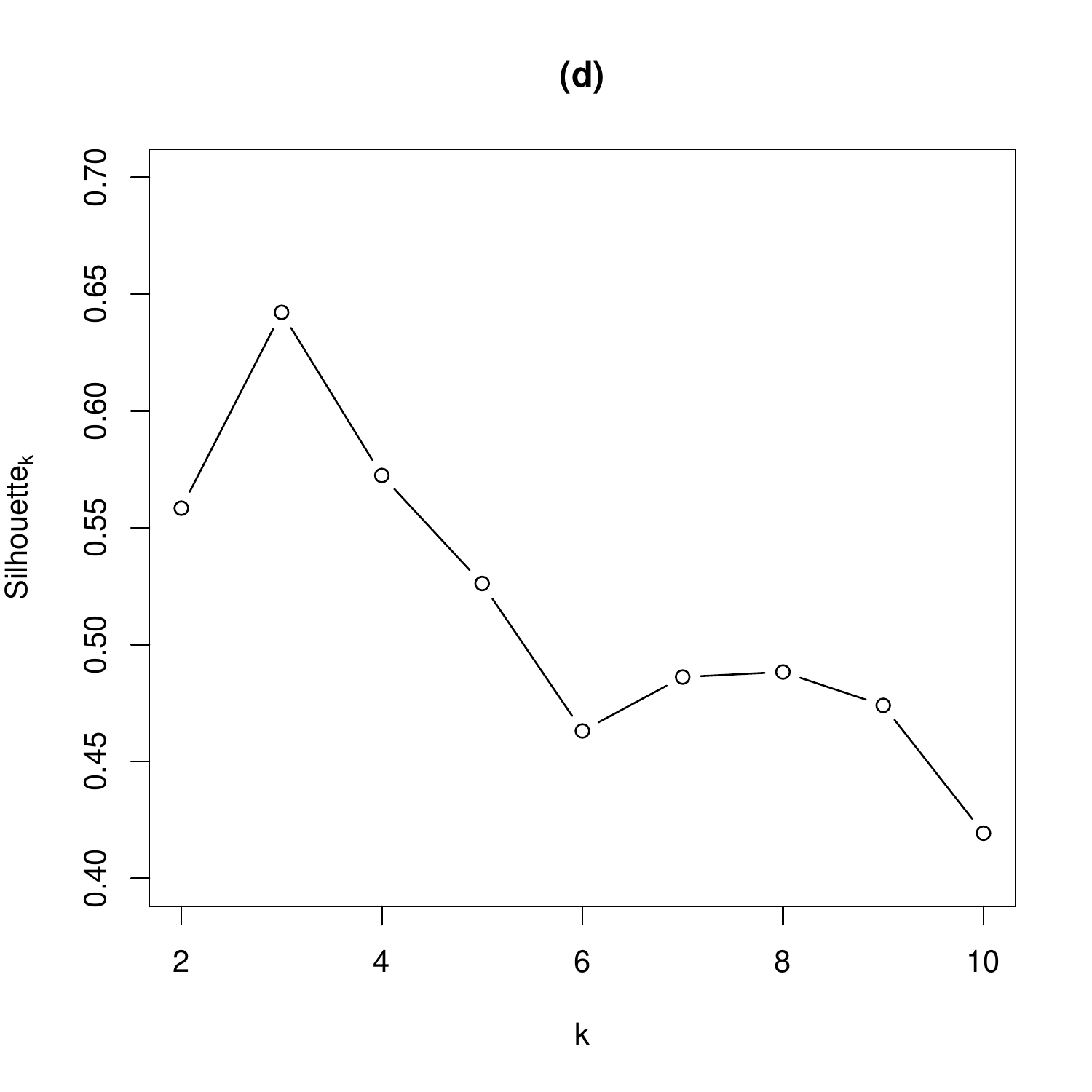}
\includegraphics[width=5.5cm]{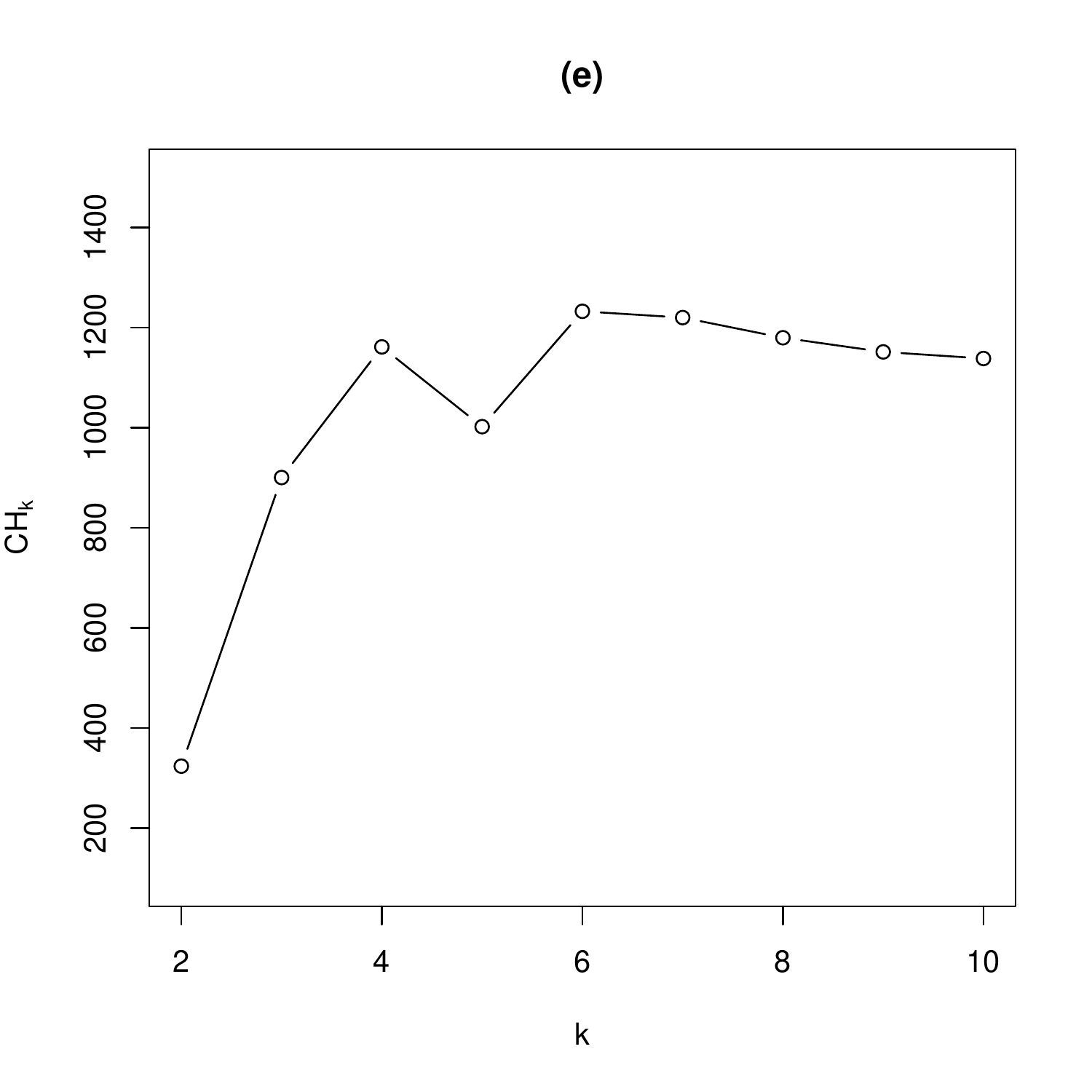}
\includegraphics[width=5.5cm]{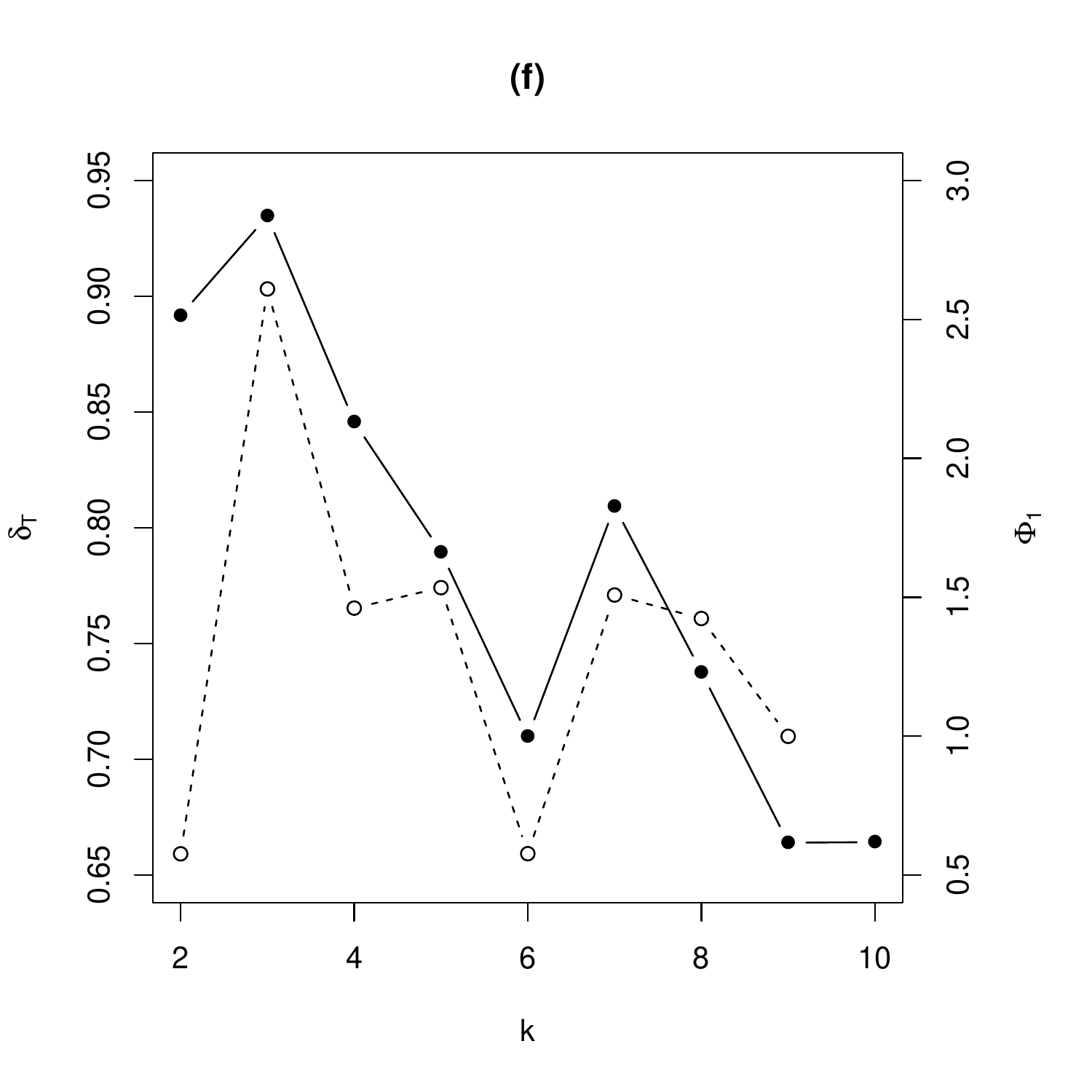}
\end{figure}
\end{landscape}}

\afterpage{
\begin{table}
\caption{The estimated $\delta_{mk}$ for the Iris dataset example given in \textbf{Example 6.1}.}
\fbox{
\begin{tabular}{c |*{4}{c}}
  Membership&  Cluster ($C_k$) \\
  ($M_m$)&  1&  2&  3\\
\hline
 1&0.985&0.004&0.011\\
 2&0.040&0.714&0.246\\
 3&0.036&0.092&0.872\\
\end{tabular}}
\end{table}

\begin{table}
\caption{\textbf{(a)} The estimated $\delta_{mk}$ for the Transcriptional \textit{Plasmodium chabaudi} Dataset given in \textbf{Example 6.2}. \textbf{(b)} The estimated $\delta_{mk}$ given in \textbf{(a)}, with a threshold upon $\delta_{mk}$ ($\delta_{mk} < 0.1$).}
\textbf{(a)}
\\
\fbox{
\begin{tabular}{c |*{5}{c}}
  Membership&  Cluster ($C_k$) \\
  ($M_m$)&  1&  2&  3& 4\\
\hline
1&0.695&0.095&0.059&0.151\\
2&0.090&0.660&0.076&0.174\\
3&0.064&0.086&0.596&0.254\\
4&0.089&0.109&0.140&0.662\\
\end{tabular}}
\\
\\
\textbf{(b)}
\\
\fbox{
\begin{tabular}{c |*{5}{c}}
  Membership&  Cluster ($C_k$) \\
  ($M_m$)&  1&  2&  3& 4\\
\hline
1&0.822&0.000&0.000&0.178\\
2&0.000&0.791&0.000&0.209\\
3&0.000&0.000&0.701&0.299\\
4&0.000&0.120&0.153&0.727\\
\end{tabular}}
\end{table}}

\subsection{Transcriptional Dataset}
A further example of applying the \textit{degree of membership} was to 23 mice and 5011 combined orthologues gene recounts for isolates CB and AS of the malaria parasite given in \cite{lin2018}. These data include both CB and AS isolates given to serially blood passaged (SBP) and mosquito-transmitted (MT) infected mice, with treatments SBP-AS ($n = 5$), MT-AS ($n = 5$), SBP-CB ($n = 5$) and MT-CB ($n = 6$). Complete linkage hierarchical clusters was used to cluster in this example.

A principal components analysis was conducted on all 5011 genes across 23 mice, the first 3 Principal Components explained 98.89\% of the variance. Principal Components 2 and 3 seem to show the most obvious separation between treatments of infected mice (Figure 2 (a-b)). Both GAP statistics estimated the optimum cluster number to be 1, suggesting no clusters over 3 Principal Components (Figure 2 (c-d)). The maximum $\phi$ ratio at lag 1 occurred at 4 clusters, with a value of 2.511 (Figure 2 (e)). The estimated \textit{degree of membership} at 4 clusters was 0.647 (Figure 2 (f)), which suggests these clusters identified within the transcriptome may not well defined over 3 Principal Components and that there is noise between clusters within the dataset. The treatment group of infected mice SBP-AS was the only group which came out as their own separate cluster (Cluster 1 $\bullet$; Figure 3). Where, $\delta_{11}$ had the highest estimated value (0.695) compared to all other clusters regarding their membership ($\delta_{22} = 0.660$, $\delta_{33} = 0.596$ and $\delta_{44} = 0.662$) (Table 2 (a)). Clusters 3 and 4 were the least defined and had the most overlap, with $\delta_{34}$ estimated as 0.254. With thresholding upon $\delta_{mk}$ of 0.01, the optimum number of clusters is still 4, where the estimated $\phi$ ratio was 3.256 and $\delta_{T}$ increases to 0.745, which suggests some clusters are well defined. However, clusters 3 and 4 are still not well separated, with $\delta_{34}$ now estimated as 0.299 (Table (2)b).

\afterpage{
\begin{landscape}
\begin{figure}
\caption{\textbf{(a)} Summary of the transcriptional data \citep{lin2018} in example \textbf{6.2} over PC1 and PC2, cluster 1 ($\circ$), cluster 2 ($\square$), cluster 3 ($\blacksquare$) and cluster 4 ($\bullet$). \textbf{(b)} Summary over PC2 and PC3. \textbf{(c)} estimated GAP, unif. \textbf{(d)} estimated GAP, PCA. \textbf{(e)} $\delta_{T}$ ($\bullet$, solid-line) and $\phi$ ratio at lag 1 ($\circ$, dashed-line). \textbf{(f)} $\delta_{T}$ ($\bullet$, solid-line) and $\phi$ ratio at lag 1 ($\circ$, dashed-line) for threshold of 0.01 upon $\delta_{mk}$.}
\includegraphics[width=5.5cm]{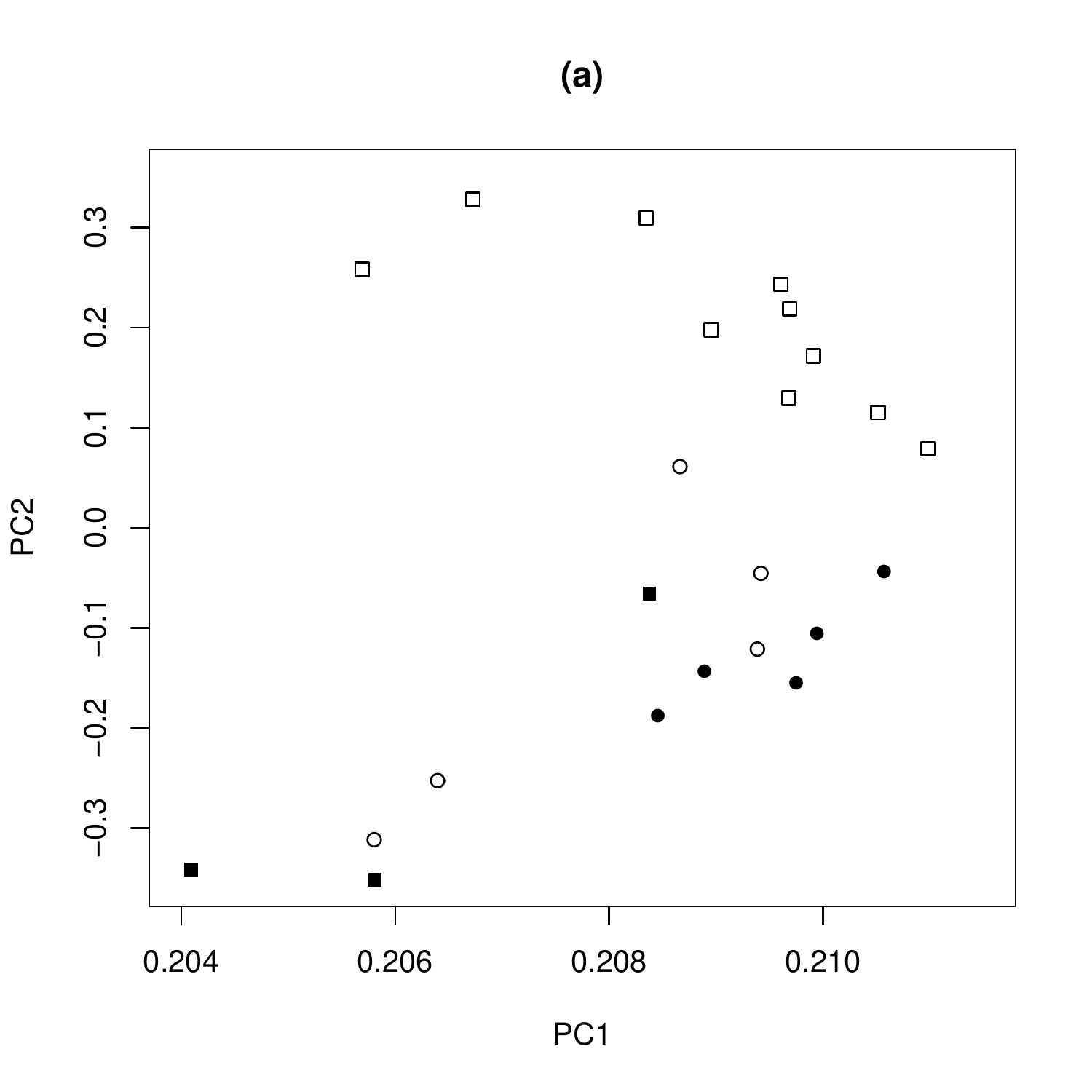}
\includegraphics[width=5.5cm]{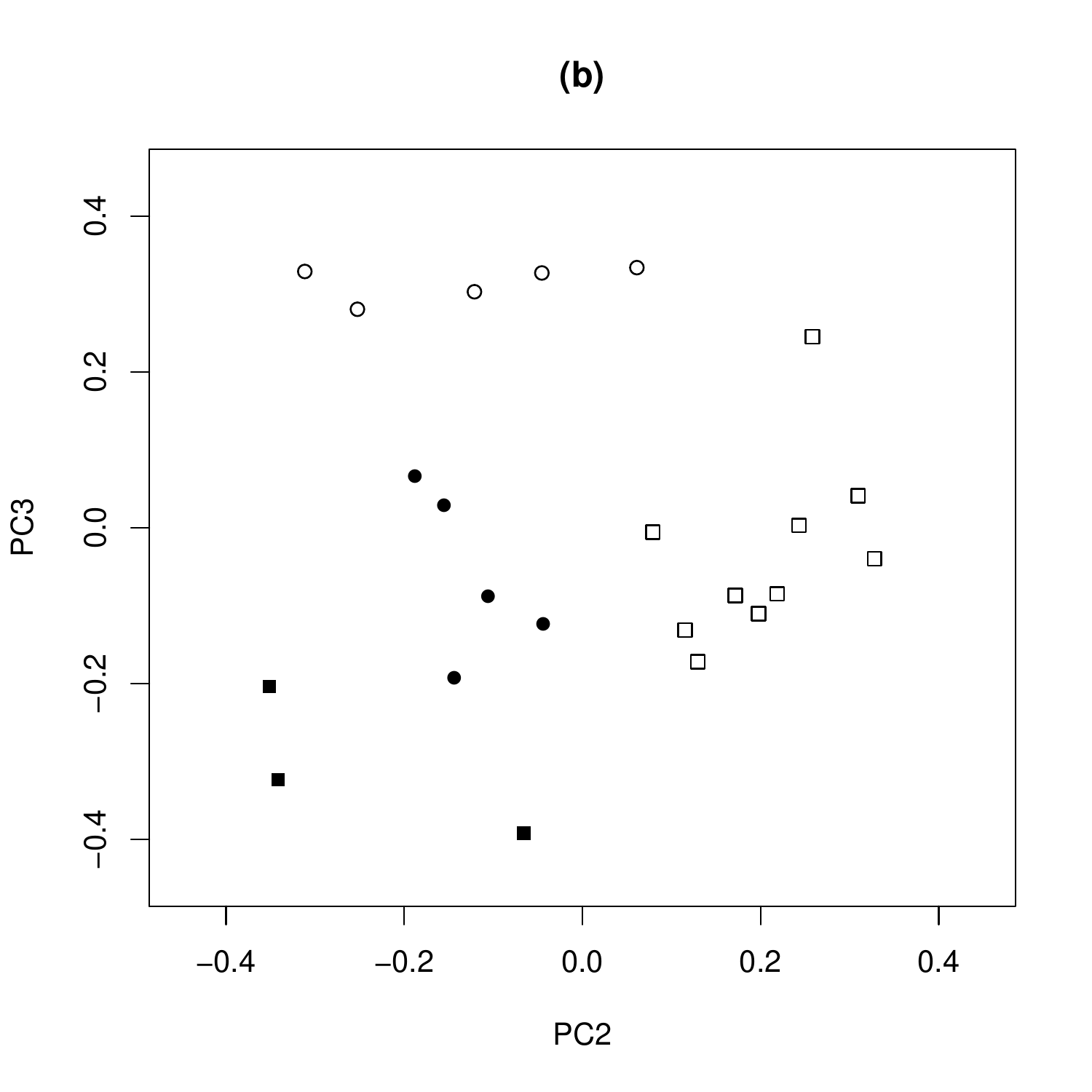}
\includegraphics[width=5.5cm]{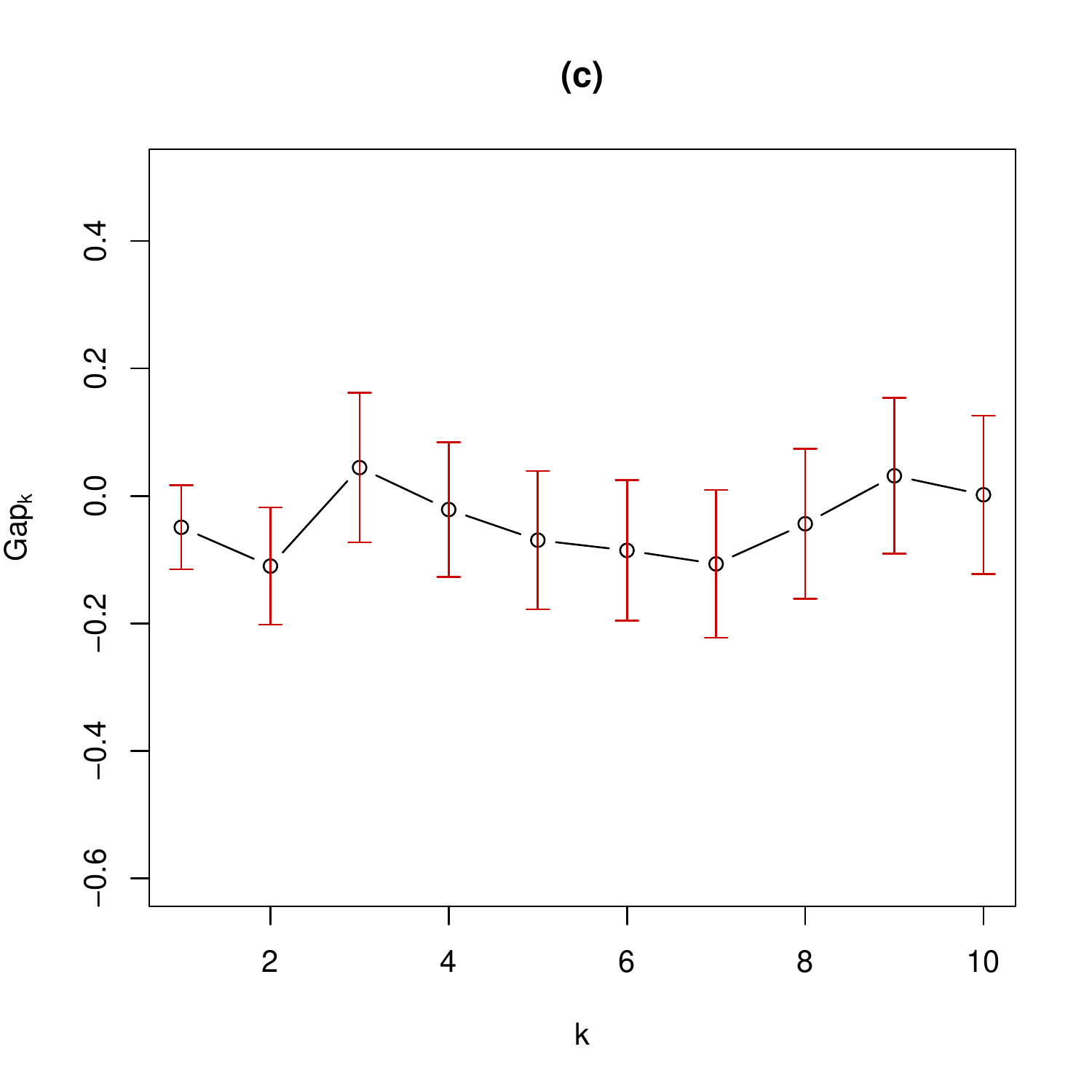}
\\
\includegraphics[width=5.5cm]{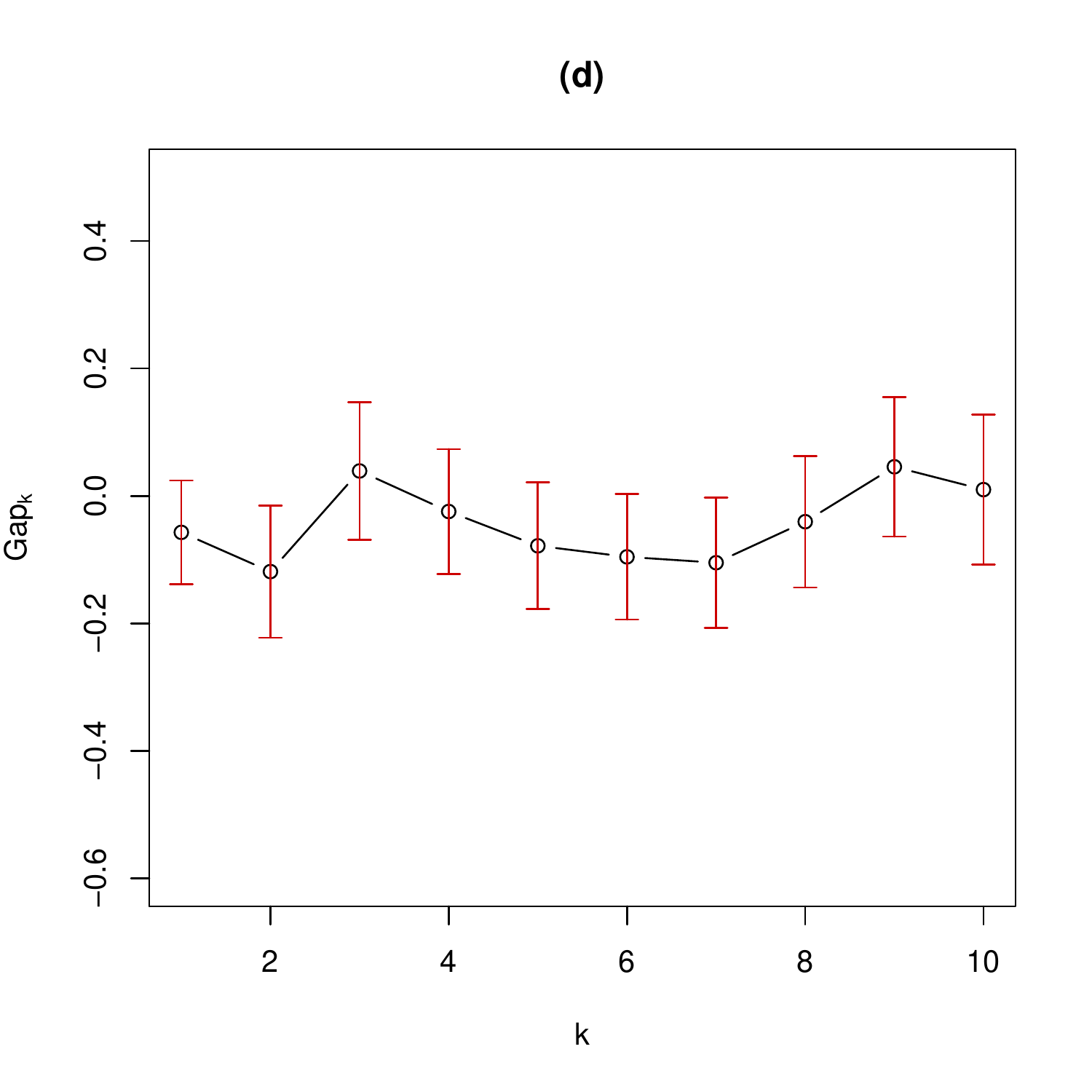}
\includegraphics[width=5.5cm]{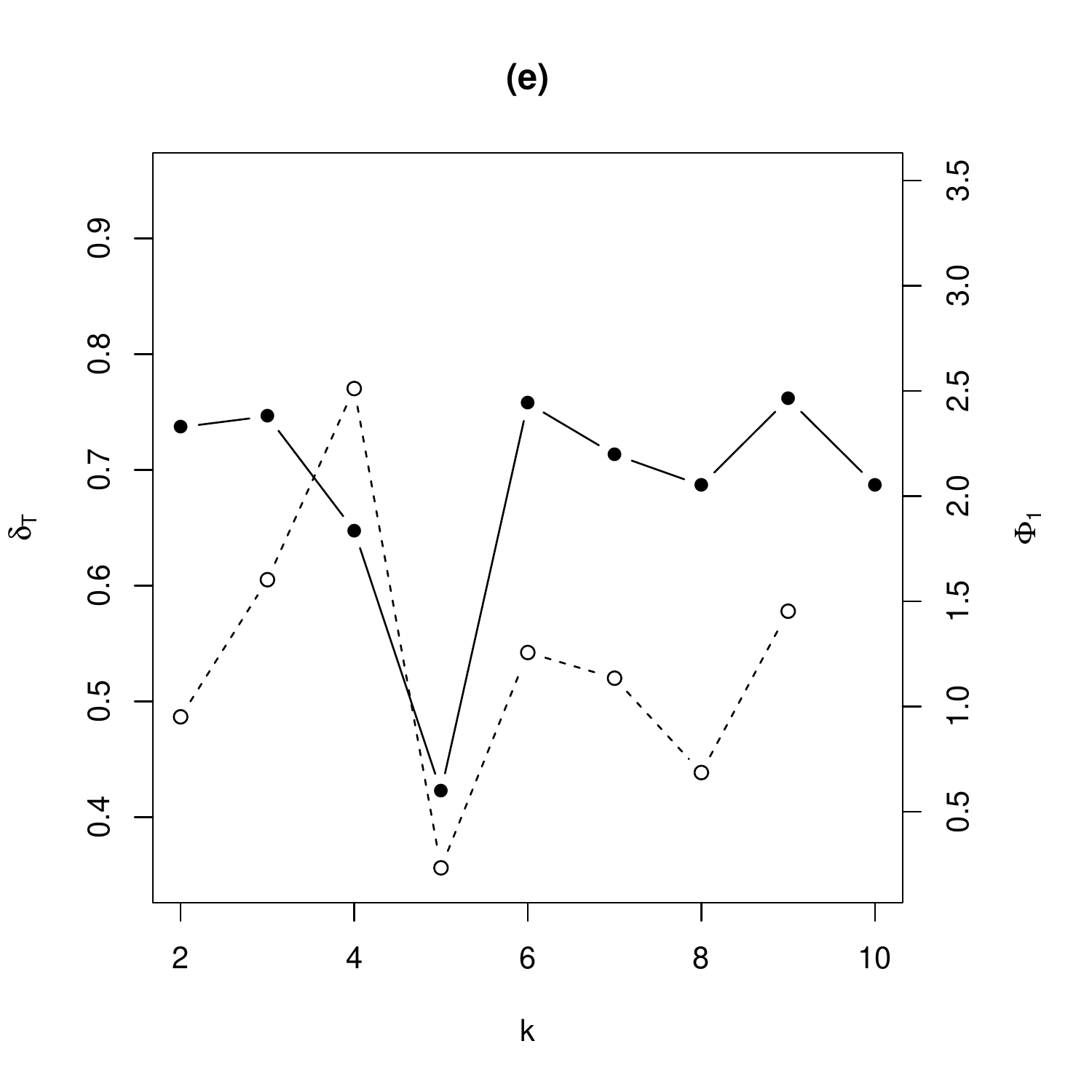}
\includegraphics[width=5.5cm]{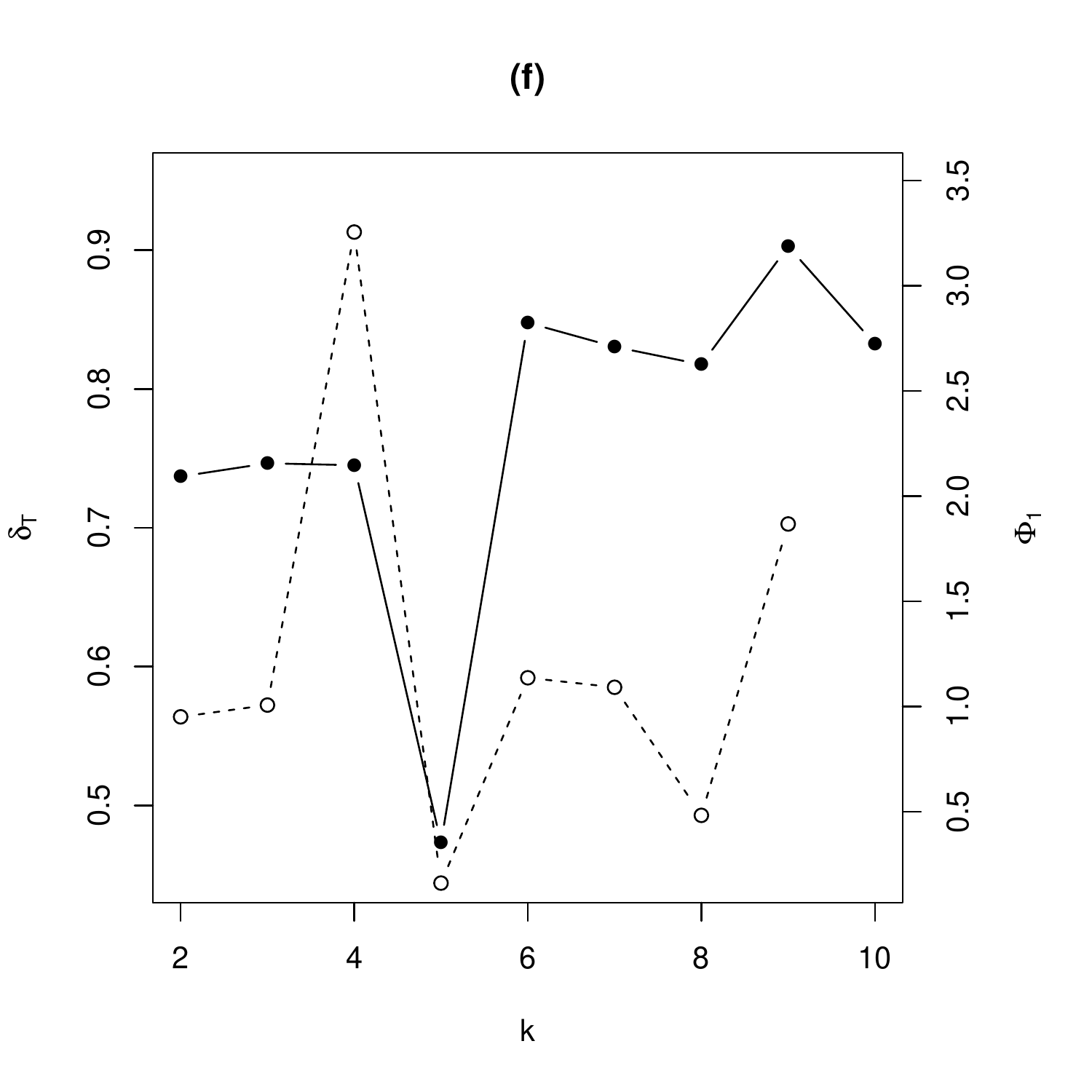}
\end{figure}
\end{landscape}}

\afterpage{
\begin{figure}
\caption{A dendrogram using complete linkage of three principal components summarising 5011 genes from 23 mice. Summary of the data and labels are given in \cite{lin2018}.}
\includegraphics[width=14cm]{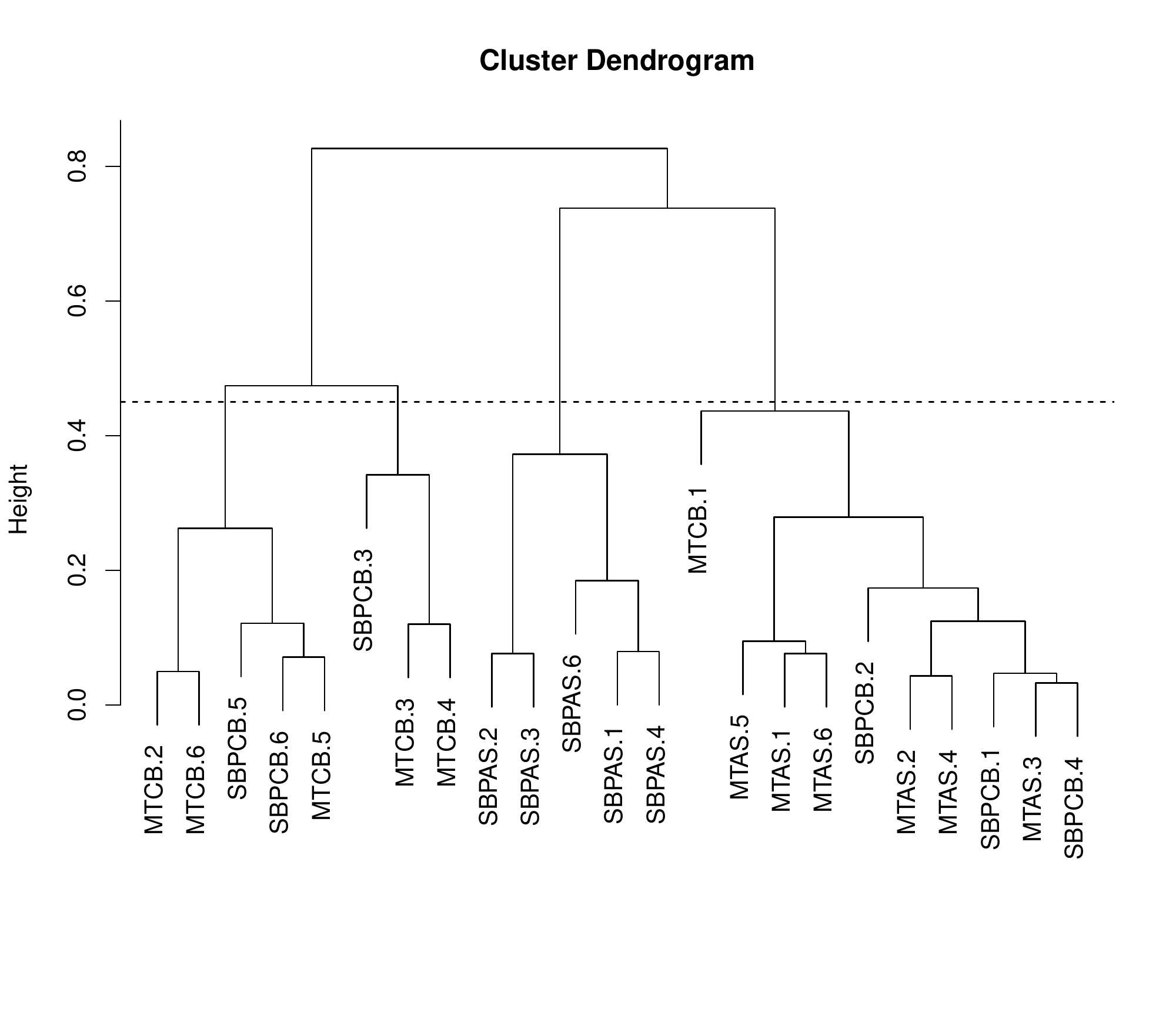}
\end{figure}}

\afterpage{
\begin{table}
\caption{The results of sampling 100 times from the simulation study in \textbf{Section 8}.}
\resizebox{!}{9.5cm}{\fbox{
\begin{tabular}{*{12}{c}}
\textit{Uniform Distribution} &  1&  2&  3&  4&  5&  6&  7&  8&  9&  10\\
\hline
GAP, unif&95&5&0&0&0&0&0&0&0&0\\
GAP, PCA&99&1&0&0&0&0&0&0&0&0\\
Silhouette&&12&5&3&3&3&9&11&18&35\\
CH&&31&17&13&9&8&7&5&7&3\\
$\phi$ ratio&&89&11&0&0&0&0&0&0&0\\
$\phi$ ratio, threshold&&89&11&0&0&0&0&0&0&0\\
\hhline{*{12}{=}}
\textit{4 clusters, 2-dim} &  1&  2&  3&  4&  5&  6&  7&  8&  9&  10\\
\hline
GAP, unif&16&0&0&84&0&0&0&0&0&0\\
GAP, PCA&18&0&0&82&0&0&0&0&0&0\\
Silhouette&&0&0&100&0&0&0&0&0&0\\
CH&&0&0&99&1&0&0&0&0&0\\
$\phi$ ratio&&1&6&88&2&1&1&1&0&0\\
$\phi$ ratio, threshold&&0&0&98&2&0&0&0&0&0\\
\hline
\textbf{N = 5}&  1&  2&  3&  4&  5&  6&  7&  8&  9&  10\\
\hline
GAP, unif&7&2&0&78&0&0&0&0&0&0\\
GAP, PCA&33&4&0&63&0&0&0&0&0&0\\
Silhouette&&0&1&99&0&0&0&0&0&0\\
CH&&0&0&11&3&1&1&3&4&77\\
$\phi$ ratio&&11&16&56&7&7&2&1&1&0\\
$\phi$ ratio, threshold&&0&0&87&7&2&1&1&2&0\\
\hhline{*{12}{=}}
\textit{4 clusters, 4-dim} &  1&  2&  3&  4&  5&  6&  7&  8&  9&  10\\
\hline
GAP, unif&0&0&0&100&0&0&0&0&0&0\\
GAP, PCA&0&0&0&100&0&0&0&0&0&0\\
Silhouette&&0&45&55&0&0&0&0&0&0\\
CH&&0&0&100&0&0&0&0&0&0\\
$\phi$ ratio&&0&1&87&1&4&4&3&0&0\\
$\phi$ ratio, threshold &&0&15&85&0&0&0&0&0&0\\
\hline
\hline
\textit{Nested Clusters} \\
\hline
\textbf{SD = 0.5}&  1&  2&  3&  4&  5&  6&  7&  8&  9&  10\\
\hline
GAP, unif&0&0&21&0&0&79&0&0&0&0\\
GAP, PCA&0&0&21&0&0&79&0&0&0&0\\
Silhouette&&0&100&0&0&0&0&0&0&0\\
CH&&0&0&0&0&94&5&1&0&0\\
$\phi$ ratio&&0&53&0&0&46&0&1&0&0\\
$\phi$ ratio, threshold&&0&49&0&0&50&0&1&0&0\\
\hline
\textbf{SD = 1}&  1&  2&  3&  4&  5&  6&  7&  8&  9&  10\\
\hline
GAP, unif&0&0&53&41&3&3&0&0&0&0\\
GAP, PCA&0&0&32&66&1&1&0&0&0&0\\
Silhouette&&0&100&0&0&0&0&0&0&0\\
CH&&0&8&8&4&66&9&3&1&1\\
$\phi$ ratio&&0&100&0&0&0&0&0&0&0\\
$\phi$ ratio, threshold&&0&100&0&0&0&0&0&0&0\\
\end{tabular}}}
\end{table}}

\section{Simulation Study}
Below is a simulation study, similar to \cite{tibshirani2001estimating} when evaluating the GAP statistic. This was applied to the following methods described in Section \textbf{5} and the $\phi$ ratio presented in Section \textbf{3}.
\begin{enumerate}[label = (\arabic*)]
\item Gap statistic using a uniform distribution \citep{tibshirani2001estimating} (Equation 6),
\item Gap statistic using PCA \citep{tibshirani2001estimating},
\item \cite{calinski1974dendrite} (Equation 7),
\item Silhouette \citep{rousseeuw1987silhouettes} (Equation 8),
\item $\phi$ ratio at lag 1 (Equation 5),
\item $\phi$ ratio at lag 1 with threshold factor upon $\delta_{mk}$ at 0.1.
\end{enumerate}
Four simulation scenarios are given and have been sampled 100 times. The results from these simulations are presented in Table 3. Complete linkage hierarchical clustering was used.
\begin{enumerate}[label = (\alph*)]
\item A \textit{uniform distribution} with 100 observations over six dimensions with a minimum of 0 and a maximum of 1.
\item \textit{Four clusters four dimensions} with centres ($-4, -6$), ($-8, 1$), ($4, -5$) and ($-6, 9$). The first simulation has a sample size of 25, 25, 10 and 10 for each cluster, the second simulation has a sample of 5 for all clusters. This was to test the adequacy of the proposed method of degree of membership with a low number of samples.
\item \textit{Four clusters in four dimensions} with centres ($11, -8, 0, -3$), ($-8, 4, 4, 2$), ($9,$ $-2,$ $-2,$ $3$), ($3$, $7,$ $-4,$ $0$) with 30, 30, 20 and 15 samples. To evaluate the bias of higher dimensional datasets.
\item \textit{Nested Clusters} with two \textit{nesting} at each cluster, 6 in total, with centres ($12, -15$), ($15, -18$), ($-16, -15$), ($-16, -18$), ($17, 14$) and ($14, 11$) with $25,$ $25,$ $15,$ $15,$ $10$ and $10$ samples. First sampled with a standard deviation of 0.5, then 1, to illustrate the effect when nested clusters (SD = 0.5) become a single cluster (SD = 1). At SD = 0.5, both 3 and 6 clusters are thought to be equally likely.
\end{enumerate}
\textit{Uniform distribution}:
Both GAP methods adequately found there were no clusters. This is due to the Silhouette, CH and $\phi$ ratio unable to be calculated for 1 cluster. Both the Silhouette and the CH both had a tendency to over cluster, whereas the $\phi$ ratio optimised around a total cluster number of 2. \\
\\
\textit{Four clusters two dimensions}: All methods found 4 to be the optimum number of clusters. The $\phi$ ratio and both GAP methods were the worst in estimating the correct number of clusters compared to all other methods. The Silhouette, CH $\phi$ ratio with threshold provided the best methods in this simulation. With fewer observations, the Silhouette and $\phi$ ratio with threshold were the better methods at estimating the correct number of clusters. The CH method became worse at estimating the correct number of clusters with fewer observations. In both cases, estimation on the optimum number of clusters through the $\phi$ ratio was improved by including a threshold at 0.1. \\
\\
\textit{Four clusters, four dimensions}: With higher dimensions, the $\phi$ ratio was similar at optimising the number of clusters at 2 dimensions. The GAP, unif, PCA and CH performed better in higher dimensions. The Silhouette method provided worst estimates in more dimensions. The $\phi$ ratio with threshold of 0.1 still provided an optimisation of 4 clusters. \\
\\
\textit{Nested clusters}: The $\phi$ ratio with threshold was the least biased when optimising the number of clusters, with a 49-50 split for 3 and 6 clusters, respectively at SD = 0.5. The Silhouette method had a bias for fewer clusters. With a more variable dataset (SD = 1), and 6 clusters becomes 3, the GAP statistics and CH methods over-performed compared to the Silhouette and $\phi$ ratio. By applying a threshold of 0.1 to the method of $\phi$ ratio for nested clusters, the method accurately identified both 3 clusters and not their nests. With increased variability (SD = 1), and therefore the disappearance of nested clusters, the $\phi$ ratio with and without a threshold and the Silhouette method outperformed the GAP statistics and the CH method.  \\
\\
There was no overall consensus on which method out-performed the others across all simulations (Table 3), as each method is suited to different types of data, where the GAP statistics are the only methods which may consider no clusters. What is consistent is that the $\phi$ ratio estimate was more consistent across simulations. With the nested simulation suiting the $\phi$ ratio best compared to other methods. Considerations of the definition of a defined cluster and threshold should also be given for different cluster scenarios. \\
\nocite{R2018}

\section{Discussion}
The use of the $\phi$ ratio at lag 1 (Equation 5) and other established clustering statistics (Equations 6-8) provided an adequate estimation of optimisation of the number of clusters for all scenarios where $k>1$. Although some have potential biases and some work better with different types of datasets. The method of $\phi$ ratio was shown to be the least bias when identifying nested clusters and with more variable datasets. Generally the $\phi$ ratio was more consistent across given scenarios. The \textit{degree of membership} provided an quantitative estimate of the homogeneity, not only for the whole cluster and membership fit, but also for each cluster. This statistic makes it possible to identify which clusters within an analysis may not be representative of their membership, or how well their membership is represented by other clusters. The Silhouette method considers a similar method but only of the neighbouring cluster. It may be the case that from a cluster analysis, more so with nested clusters, that an optimum number of clusters may be found, however, the membership of closely related clusters may blur the accuracy of identified clusters. It is proposed the heterogeneity may be evaluated for each cluster and each membership through $\delta_{mk}$.\\
\\
The method of optimising cluster number through the $\phi$ ratio was improved by introducing a threshold upon $\delta_{mk^\neg}$ for defined clusters. Although cluster analysis may not be an application-independent mathematical problem \citep{von2012clustering}, where prior experimental knowledge may explain clusters, each cluster and its membership within one analysis is not application-independent, such that some clusters may be more biologically relevant and obvious in their membership, while others may be more varied and so a consideration of a threshold upon $\delta_{mk}$ should be considered. However, considerations on the threshold should be given as this may vary depending on the dimensionality and if any nested clusters are present within the dataset. Overall the use of the same threshold for all cluster analyses was shown to be inadequate and varied between simulations. Therefore, similar to the application of cluster analysis \citep{von2012clustering}, the use of a threshold upon $\delta_{mk}$ is not application-independent and is dependent on the relevance of the applied problem and leads to more of a problem-centric perspective.\\
\\
The application of classifying a dataset into optimal meaningful clusters is a difficult process dependent on the sampled data, method of clustering and method of optimising used. This paper presents that considerations of the heterogeneity of each cluster through cluster membership should be acknowledged. The method gives consideration of how membership of a cluster should also be considered along with the cluster fit itself.

\section{Acknowledgements} 
A special thanks to JJ Valletta, K Hassall, V Koutra, H Metcalf and G Kelly for reading an early draft of this manuscript and providing valuable feedback.
\\
\\
The Francis Crick Institute receives its core funding from Wellcome (FC001101), the UK Medical Research Council (FC001101) and Cancer Research UK (FC001101). This study was funded by the UK Medical Research Council (MR/M003906/1). Jean Langhorne is a Wellcome Trust Senior investigator (102907).

 \bibliography{BibforHCPaper}
 \bibliographystyle{agsm}

\end{document}